\documentclass{nature}

\usepackage[utf8]{inputenc}
\usepackage{hyperref}
\usepackage{graphicx}
\usepackage{siunitx}
\usepackage[T1]{fontenc}

\usepackage[font={small}, labelfont={small, bf}]{caption}

\makeatletter
\let\saved@includegraphics\includegraphics
\AtBeginDocument{\let\includegraphics\saved@includegraphics}
\renewenvironment*{figure}{\@float{figure}}{\end@float}
\makeatother

\title{Dynamic decoupling of laser phase noise in compound atomic clocks}

\author{Sören Dörscher$^1$, Ali Al-Masoudi$^{1, 4}$, Marcin Bober$^2$, Roman Schwarz$^1$, Richard Hobson$^3$, Uwe Sterr$^1$ \& Christian Lisdat$^1$}

\date{\today}

\begin{document}
\maketitle
\begin{affiliations}
 \item Physikalisch-Technische Bundesanstalt, Bundesallee 100, 38116 Braunschweig, Germany
 \item Institute of Physics, Faculty of Physics, Astronomy and Informatics, Nicolaus Copernicus University, Grudziądzka 5, 87-100 Toruń, Poland
 \item National Physical Laboratory, Teddington, Middlesex TW11 0LW, United Kingdom
 \item Current address: IAV GmbH, Nordhoffstraße 5, 38518 Gifhorn, Germany
\end{affiliations}

\begin{abstract}
The frequency stability achieved by an optical atomic clock ultimately depends on the coherence of its local oscillator.
Even the best ultrastable lasers only allow interrogation times of a few seconds, at present.
Here we present a universal measurement protocol that overcomes this limitation.
Engineered dynamic decoupling of laser phase noise allows any optical atomic clock with high signal-to-noise ratio in a single interrogation to reconstruct the laser's phase well beyond its coherence limit.
A compound clock is then formed in combination with another optical clock of any type, allowing the latter to achieve significantly higher frequency stability than on its own.
We demonstrate implementation of the protocol in a realistic proof-of-principle experiment with a phase reconstruction fidelity of \SI{99}{\percent}.
The protocol enables minute-long interrogation for the best ultrastable laser systems.
Likewise, it can improve clock performance where less stable local oscillators are used, such as in transortable systems.

\end{abstract}
\section*{\label{sec:introduction}Introduction}
The progress of optical clocks has enabled a multitude of applications that range from testing fundamental symmetries underlying relativity\cite{san19, del17} and searching for physics beyond the standard model,\cite{ros08, hun14, god14} including dark matter,\cite{der14, wci16, sta16c, wci18a, rob19b} to measuring geopotential differences\cite{gro18a} and the proposed use for gravitational wave detection.\cite{kol16}
Since lower frequency instability of a clock reduces the time required to perform measurements with a given precision, it benefits applications in general and those where time-dependent effects are measured, including transient changes of fundamental constants,\cite{rob19b} in particular.
Therefore, the advancement of ultrastable lasers\cite{hae15a, mat17a} and other techniques to improve measurement instability\cite{cho11, tak11, hum16, sch17, oel19} continue to be a focus of research.

The frequency stability of optical clocks is limited by quantum projection noise\cite{ita93} (QPN) as well as noise resulting from non-continuous observation of the laser frequency, which is known as the Dick effect.\cite{dic87, que03}
The contribution of the latter depends intricately on the noise spectrum of the laser and the parameters of clock operation, but can be minimised by using Ramsey spectroscopy with little or no dead time.
In this case, QPN limits the instability of an atomic clock, given by the Allan deviation $\sigma_y$, to the standard quantum limit (SQL)
\begin{equation}
\sigma_y (\tau) = \frac{1}{2 \pi \nu_0}\sqrt{\frac{T_\mathrm{c}}{N T_\mathrm{i}^2 \tau}}
	\label{eq:sql}
\end{equation}
if $N$ uncorrelated atoms or ions are interrogated, where $T_\mathrm{i}$ is the interrogation time, $T_\mathrm{c}$ the total cycle time, and $\tau$ the averaging time.\cite{rie04a}

A clock’s frequency instability is thus minimised by using the longest possible interrogation time.
It is, in fact, the only way to improve the frequency instability for single-ion clocks, which are limited by their significant QPN ($N=1$) through Eq.~(\ref{eq:sql}).
The situation is more complex in optical lattice clocks, which benefit from their lower projection noise ($N \gg 1$).
Their frequency instability can be improved by reducing projection noise, including the use of spin squeezing,\cite{win92} and by rejection of aliased laser frequency noise using techniques such as synchronous\cite{tak11} or dead time–free interrogation.\cite{sch17, oel19}
These are complementary to maximising interrogation time and can be combined to achieve the best possible frequency stability.

However, the coherence of even the most stable lasers\cite{mat17a} limits interrogation times well short of the excited states’ natural lifetimes for typical clock transitions.
For Ramsey interrogation, the signal depends sinusoidally on the phase difference $\Delta\phi$ that accumulates between the laser light field and the atomic oscillator during interrogation.
Thus, phase deviations due to laser noise can only be traced unambiguously within a single fringe ($\left|\Delta\phi\right| < \pi / 2$).
The coherence time $T_\mathrm{co}$  can then be defined by demanding that this threshold be exceeded in no more than \SI{1}{\percent} of all cases.\cite{mat17a}\footnote{Here we have corrected the definition found in ref.~\citenum{mat17a} by reducing the phase threshold, and thus the laser coherence time, by a factor of two.}
This corresponds to a coherence time of about $T_\mathrm{co} = \SI{5.5}{\second}$ at a frequency of \SI{194}{\tera\hertz} for the best state-of-the-art lasers,\cite{mat17a} whereas typical excited-state lifetimes of clock transitions range from \SI{20.6}{\second} for $\mathrm{Al}^{+}$ (ref.~\citenum{ros07}) to several years for the electric octupole transition in $\mathrm{Yb}^{+}$ (ref.~\citenum{rob97}).

\begin{figure}
	\centerline{\includegraphics{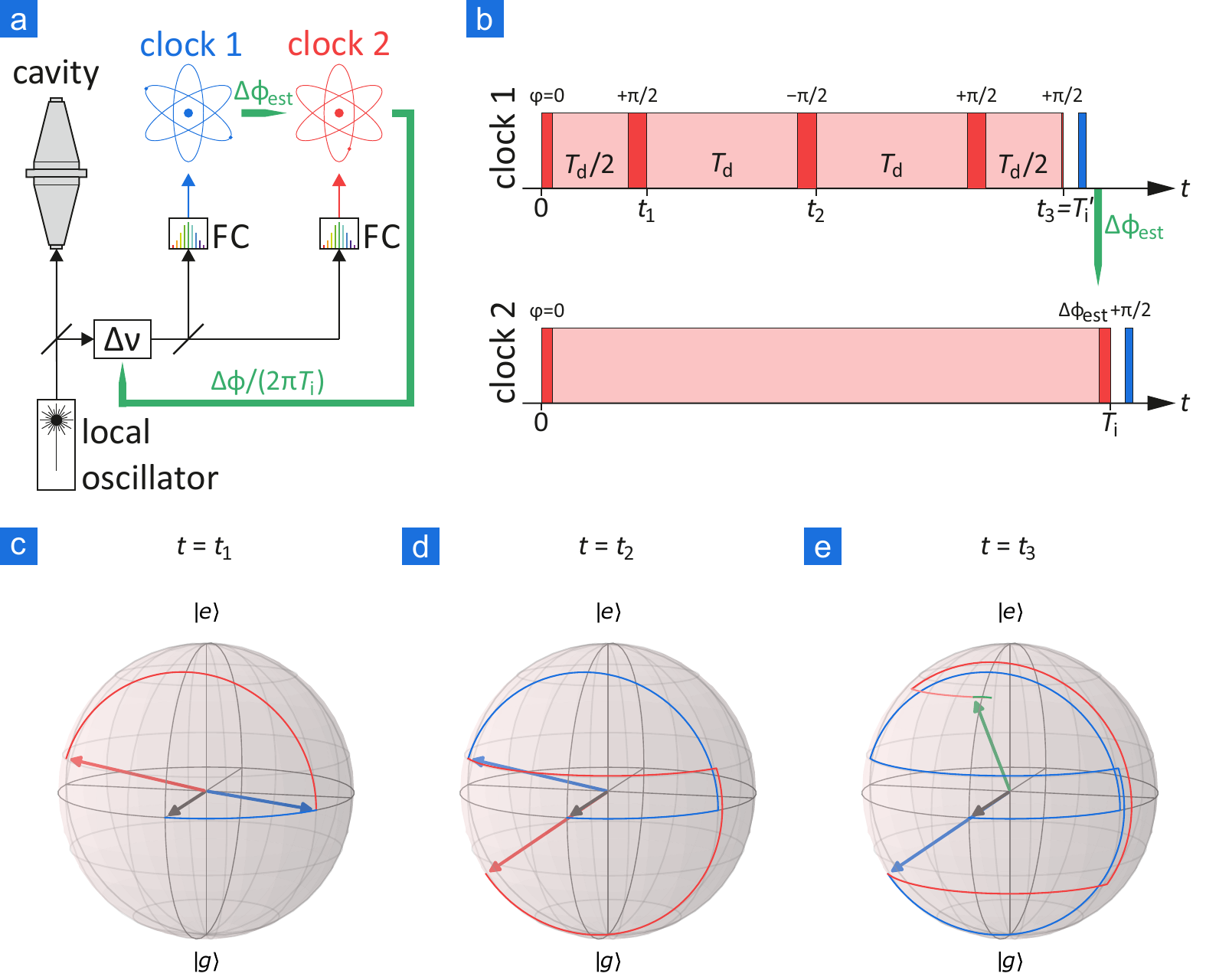}}
	\caption{
	{\textbf{(a)}}
	Schematic setup of a compound clock for operation beyond the laser coherence limit.
	The two clocks share a common local oscillator (LO) that is pre-stabilised to an ultrastable cavity.
	The frequency stability of the LO is transferred to interrogation lasers, e.g., by a frequency comb (FC).
	Using the spectroscopic sequence shown in panel b, clock 1 provides a coarse estimate of the laser phase deviation ($\Delta\phi_\mathrm{est}$) to clock 2, which refines the measurement ($\delta\phi$). The measured phase deviation $\Delta\phi = \Delta\phi_\mathrm{est} + \delta\phi$ is used to feed back to a frequency shifter ($\Delta\nu$) and stabilize the LO frequency.
	Note that the measured phase and frequency deviations need to be scaled by the frequency ratio when transferred to a clock or LO operating at a different frequency, which has been omitted here for the sake of simplicity.
	{\textbf{(b)}}
	Pulse sequences of clocks 1 and 2 (example).
	After an initial $\pi/2$ excitation pulse, the interrogation sequence of clock 1 interleaves free-evolution times of duration $T_\mathrm{d}$ or $T_\mathrm{d}/2$ (light red) and ‘flip’ pulses (red) of pulse area $\pi - \epsilon$ and phase $\varphi = \pm\pi/2$ with respect to the initial pulse.
	It ends with a pulse of area $\epsilon / 2$ and state read-out (blue).
	Clock 2 uses a two-pulse Ramsey sequence.
	It receives laser phase information ($\Delta\phi_\mathrm{est}$) from clock 1 in time to adjust the phase of the second $\pi / 2$ pulse such that the fringe center is shifted to maximise the signal slope.
	The delay $T_\mathrm{i} - T_\mathrm{i}^\prime$ must be kept short to avoid excess phase noise (see Methods).
	{\textbf{(c–-e)}}
	Evolution of the atomic state in clock 1 on the Bloch sphere for constant laser detuning (example).
	After accumulating phase during the first dark time $T_\mathrm{d} / 2$ (panel c, blue), a flip pulse nearly reverses this precession of the Bloch vector and maps it onto a small change of excitation probability (panel c, red).
	The process is repeated twice with dark time $T_\mathrm{d}$  (panels d and e, red).
	Finally, another free-evolution time $T_\mathrm{d} / 2$ (panel e, light red) and the last laser pulse with area $\epsilon / 2$ (panel e, green) are applied.
	\label{fig:protocol}
	}
\end{figure}
Operating a clock beyond this laser coherence limit is highly desirable for enhanced frequency stability.
Dynamic decoupling methods\cite{vio99} can prevent decoherence from a variety of noise sources, e.g., spin echo methods suppress inhomogeneous broadening.\cite{hah50}
However, operating a clock is tantamount to measuring the average laser frequency with respect to the atomic resonance;
thus perfect decoupling of laser frequency noise is not useful.
Here, we present a coherent multi-pulse interrogation scheme that partially decouples laser noise in a well-controlled fashion (see Fig.~\ref{fig:protocol}), with similarities to pulse-echo sequences.
It retains enough information to derive an estimate of phase deviations $\Delta\phi$ over the full interrogation time $T_\mathrm{i}$, which can easily exceed the laser coherence time.
This information can then be combined with regular Ramsey interrogation by a second clock (see Fig.~\ref{fig:protocol}) to unambiguously and precisely determine $\Delta\phi$ even well beyond $\pm\pi / 2$.
The frequency stability of this compound clock is thus superior to a single apparatus with Ramsey interrogation of duration $T_\mathrm{i} \le T_\mathrm{co}$ (see Eq.~(\ref{eq:sql})).
\section*{Results}
\begin{figure}
	\centerline{\includegraphics{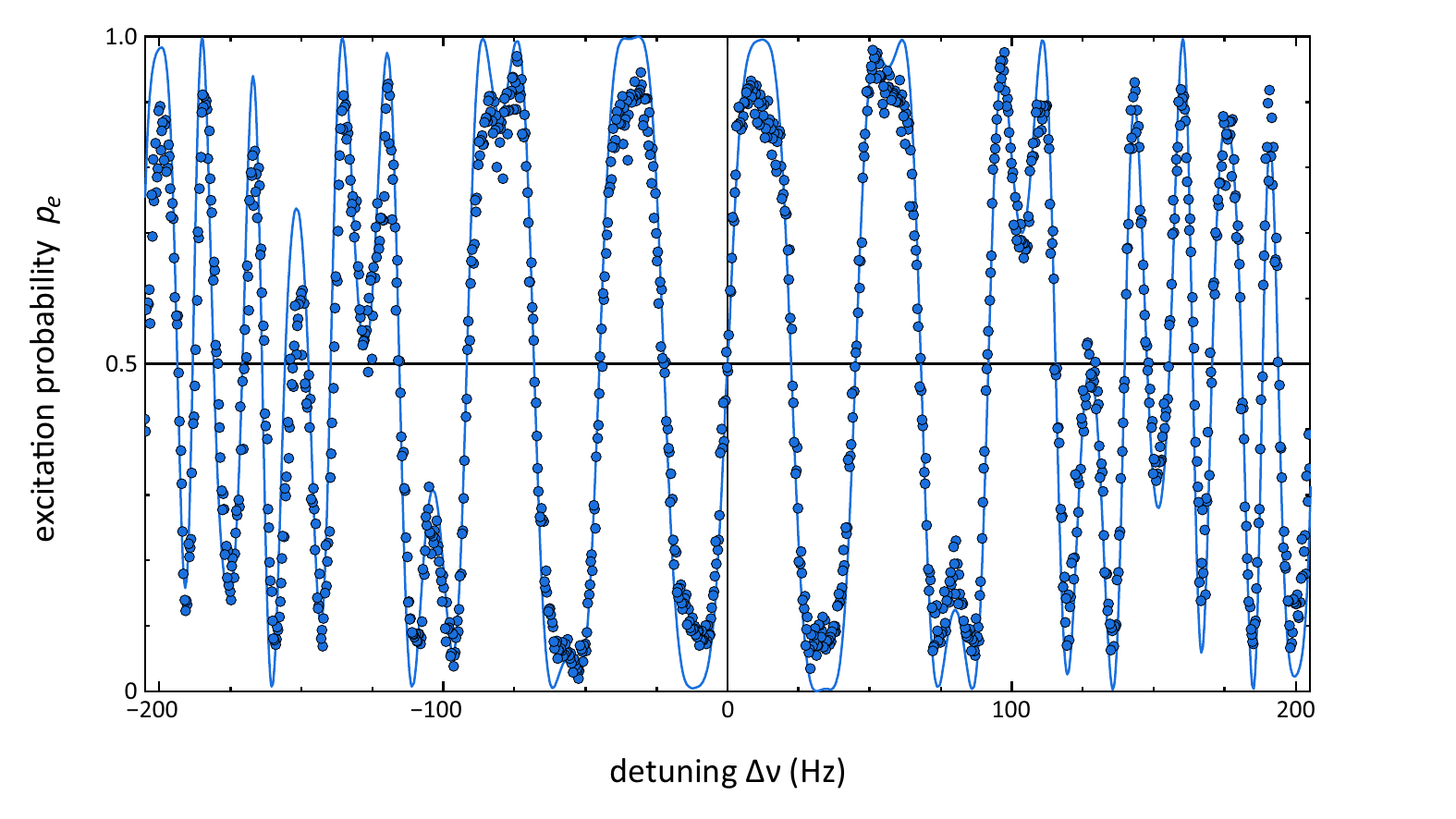}}
	\caption{
		Spectroscopy signal of clock 1 as a function of laser detuning (for $M = 6$, $\epsilon = 0.08 \pi$, $T_\mathrm{d} = \SI{40}{\milli\second}$, and a $\pi$-pulse duration $T_\pi = \SI{1}{\milli\second}$).
		The line shape expected from theory in the absence of frequency noise (solid line) features a broad central fringe as well as a complex structure far off resonance.
		Measurements using our strontium lattice clock without artificial noise (circles) reproduce this line shape very well.
		Deviations can be attributed to limited contrast (around \SI{90}{\percent}) and fluctuations of laser intensity.
		The central fringe covers about \SI{25}{\hertz}, whereas the central fringe of Ramsey spectroscopy covers only about \SI{2}{\hertz} at the same interrogation time ($T_\mathrm{i} \approx \SI{0.25}{\second}$).
		\label{fig:lineshape}
	}
\end{figure}
The multi-pulse spectroscopy protocol partially decouples clock 1 from laser noise by applying a variable number $M$ of ‘flip’ pulses that divide the free evolution time into short periods of durations $T_\mathrm{d}$  or $T_\mathrm{d} / 2$ (see Fig.~\ref{fig:protocol} for details).
Each pulse maps the phase accumulated during free evolution onto atomic state population;
hence $T_\mathrm{d}$ is the relevant timescale for laser decoherence.
The protocol allows tailoring its specific implementation to the laser noise spectrum and interrogation time, as discussed below.
An example of the resulting line shape is shown in Fig.~\ref{fig:lineshape}.

We perform a proof-of-principle experiment to demonstrate this scheme, using a strontium optical lattice clock\cite{gre16} and a state-of-the-art interrogation laser system\cite{mat17a} (see Methods).
Artificial frequency noise is imprinted onto the laser frequency to reduce the laser coherence time, which allows characterising the fidelity of phase reconstruction.
The excellent frequency stability of our interrogation laser as well as the ability to reproduce the artificial frequency noise at will (see Methods) allow us to demonstrate operation of a compound clock using sequential interrogation of a single clock.

\begin{figure}
	\centerline{\includegraphics{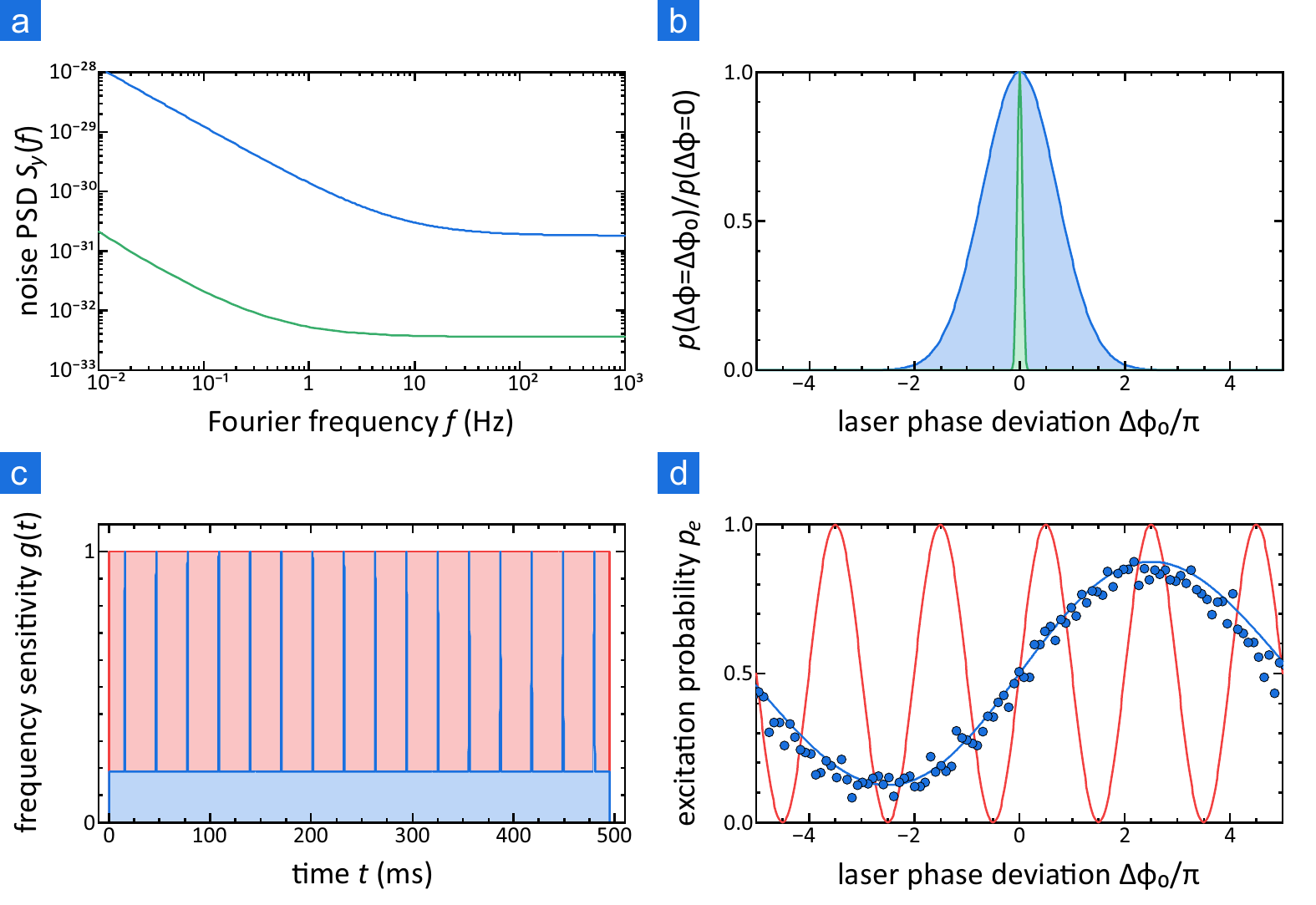}}
	\caption{
		Summary of the experimental setup for the compound clock demonstration (see Fig.~\ref{fig:exp_results}):
		{\textbf{(a)}}
		Single-sided power spectral density (PSD) $S_y(f)$ of the artificial laser noise imprinted onto the interrogation laser (blue solid line).
		The phase noise of our interrogation laser system as reported in ref.~\citenum{mat17a} is shown schematically for comparison (green line).
		{\textbf{(b)}}
		Distribution of final phase deviations due to artificial frequency noise (blue) and intrinsic laser noise (green) after the interrogation time.
		{\textbf{(c)}}
		Frequency sensitivity of regular Ramsey spectroscopy\cite{dic87, que03} (red line) and of the decoupled protocol (blue line).
		{\textbf{(d)}}
		Line shape of Ramsey spectroscopy (red line) and the decoupled protocol (blue line).
		The line shape observed when scanning the laser across resonance in the absence of artificial frequency noise is shown for comparison (blue circles).
		The contrast of the decoupled protocol’s line shape has been adjusted to match this scan.
		\label{fig:exp_scenario}
	}
\end{figure}
The experiment is summarised in Fig.~\ref{fig:exp_scenario}.
We use a noise spectrum (see Fig.~\ref{fig:exp_scenario}a and Methods) with a coherence time $T_\mathrm{co} = \SI{77}{\milli\second}$ for Ramsey interrogation and an interrogation time $T_\mathrm{i} \approx \SI{495}{\milli\second}$.
Under these conditions, the standard deviation of the laser phase well exceeds $\pi / 2$;
the phase deviations used in our experiment span five Ramsey fringes in total (see Fig.~\ref{fig:exp_scenario}b).

\begin{figure}
	\centerline{\includegraphics{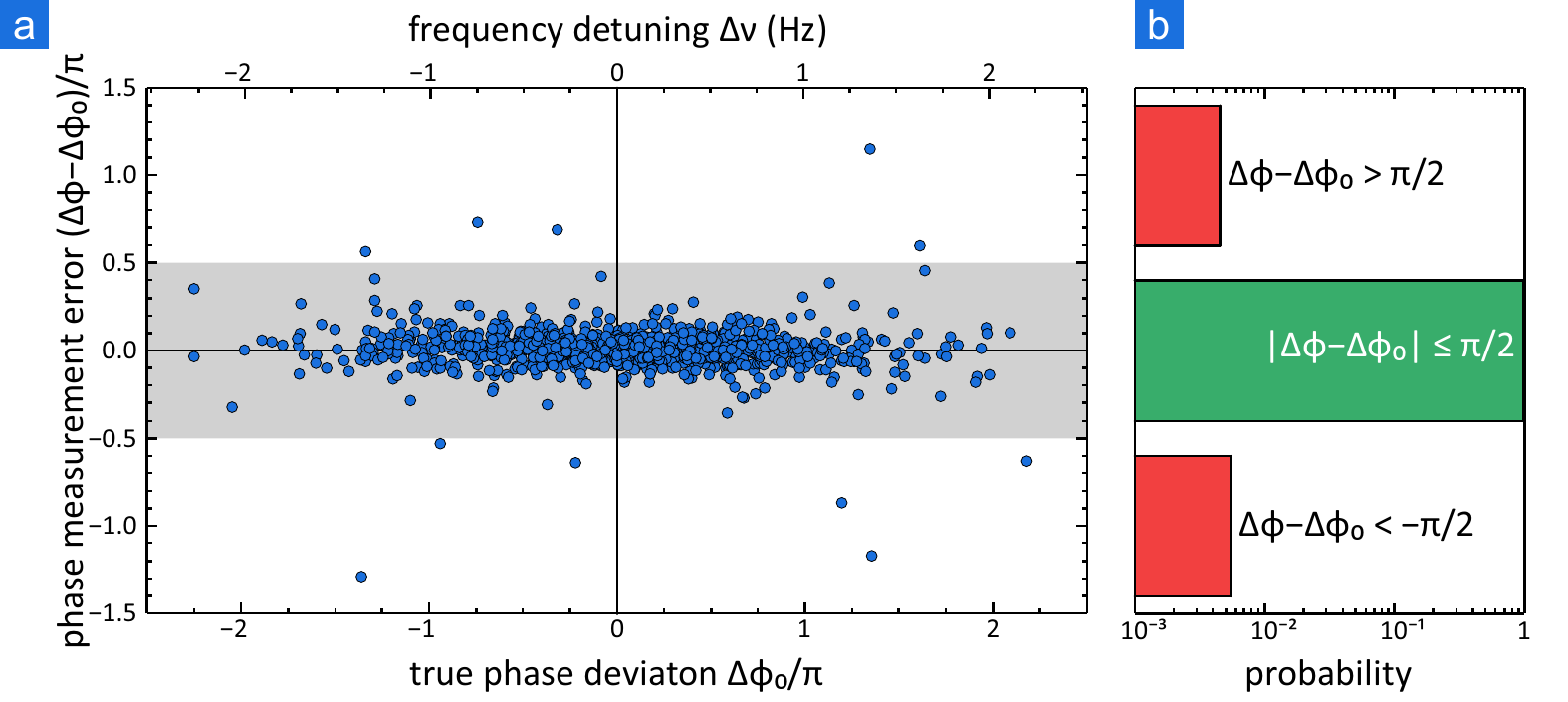}}
	\caption{
		Results of the compound clock demonstration.
		{\textbf{(a)}}
		Phase measurement errors (blue circles) for 1093 unique samples of artificial frequency noise (see Fig.~\ref{fig:exp_scenario}), where $\Delta\phi_0$ is the true phase excursion determined for each sample.
		Reconstruction is considered valid if the absolute error is $\pi / 2$ or less (grey area).
		{\textbf{(b)}}
		Probability of occurence $p$ of valid (green) and invalid (red) phase reconstruction in panel (a).
		The fidelity of this demonstration is $\mathcal{F}  \equiv p(\left|\Delta\phi - \Delta\phi_0\right| \le \pi / 2 ) = \num{0.990(3)}$.
		\label{fig:exp_results}
	}
\end{figure}
Figure \ref{fig:exp_results} shows the results of phase reconstruction of more than one thousand unique noise samples by the compound clock.
Reconstruction works well across the entire range of phase deviations.
Similar to the coherence limit for Ramsey spectroscopy, we quantify the fidelity of the compound clock using a threshold of $\pi / 2$ for the phase error (see Fig.~\ref{fig:exp_results}b).
We find an excellent fidelity $\mathcal{F} = \num{0.990(3)}$.
Furthermore, we have repeated the experiment using the same parameters for spectroscopy but increasing the added flicker frequency noise by one half, which reduces the coherence time to $T_\mathrm{co} = \SI{56}{\milli\second}$;
we still find a high, if slightly reduced fidelity $\mathcal{F}^\prime = \num{0.931(8)}$. 
This proof-of-principle experiment thus demonstrates that a compound clock using the spectroscopy scheme presented here can be interrogated well beyond the laser coherence time ($T_\mathrm{i} / T_\mathrm{co} \approx 6.4$) with a fidelity of \SI{99}{\percent}.
\section*{Discussion}
The implementation of the decoupling protocol in clock 1 is tailored to the specific laser noise spectrum of the LO and to the interrogation time.
Using the number of flip pulses $M$ and pulse defect $\epsilon$, decoupling is adjusted such that the central fringe of the resulting line shape covers most of the laser’s phase excursions.
Longer interrogation times $T_\mathrm{i}$ require a larger number of flip pulses $M$, to keep the dark time $T_\mathrm{d} \approx T_\mathrm{i} / M$ below the laser coherence limit.

QPN and imperfections of the dynamic decoupling give rise to noise of the phase estimate $\Delta\phi_\mathrm{est}$ in clock 1.
The compound clock scheme presented here can only be effective if the uncertainty of this phase estimate remains below $\pi / 2$.
A reduction of phase estimation noise in clock 1 thus allows longer interrogation of the clock transition.
This makes clocks with intrinsically low QPN, such as optical lattice clocks, ideal candidates for phase estimation (clock 1).
In contrast, clock 2 may well be a clock with low signal-to-noise ratio, e.g., a single-ion clock.

Imperfections of the dynamic decoupling result from increased frequency sensitivity during the flip pulses, as shown in Fig.~\ref{fig:exp_scenario}c.
Similar to the well-known Dick effect,\cite{dic87, que03} they result in measurement noise due aliasing of laser frequency noise.
The effect can be minimised by reducing the number of flip pulses $M$ or by decreasing their duration (see Methods for details).

Therefore, limitations of the maximum interrogation time $T_\mathrm{i}$ are of technical nature, in practice, such as the QPN of clock 1 or available interrogation laser power.
However, the requirements for minute-long interrogation are moderate:
about $10^4$ atoms and a $\pi$-pulse duration $T_\pi \approx \SI{1}{\milli\second}$ are sufficient for the case of a strontium lattice clock as clock 1 and the laser system reported in ref.~\citenum{mat17a} as LO.
For comparison, this laser system has a thermal noise floor of about \num{4E-17}, which results in a coherence time $T_\mathrm{co} \approx \SI{2.5}{\second}$ for Ramsey interrogation at the strontium clock's frequency;
minute-long Ramsey interrogation would require much lower fractional frequency instability of the laser system, $\sigma_y (\tau = T_\mathrm{co}) \approx 2 \times 10^{-18}$ at $T_\mathrm{co} = \SI{60}{\second}$.

Frequency instability of the compound clock improves according to Eq.~(\ref{eq:sql}) if clock 2 is limited by QPN (e.g., single-ion clocks) and by similar factors even if it is limited by the Dick effect (e.g., optical lattice clocks) for typical ultrastable lasers.
The scheme presented in this Letter thus reduces the duration of high-performance measurements with optical clocks by an order of magnitude and more.

\section*{Conclusion}
Improving the frequency stability of atomic clocks greatly benefits their applications, and it expedites their evaluation.
The search for or monitoring of time-dependent effects, in particular, depends critically on this aspect of clock performance.
Recent examples include tests of Lorentz symmetry\cite{san19} or the search for ultralight scalar dark matter using optical clocks.\cite{rob19b}
Many of these applications rely on ion clocks because they are especially sensitive to the effects being investigated.
Our scheme is particularly well suited in these cases, as it allows single-ion clocks to achieve frequency instability of few $10^{-17} (\tau/\si{\second})^{-1/2}$ according to Eq.~(\ref{eq:sql}).

Compared to some other proposed schemes to operate clocks beyond the laser coherence limit,\cite{cho11, hum16} the one presented here can be used for a wide variety of optical clocks of different types and frequencies.
Moreover, it enhances the absolute frequency stability of clock 2, i.e., clock performance improves in any measurement rather than only in correlated comparisons of clocks.
This sets it apart from coherent interrogation of two clocks as proposed in ref.~\citenum{cho11}, which relies on common-mode rejection of laser phase noise.
However, the phase estimation protocol improves the performance even of such correlated measurements, e.g., the proposed detection of gravitational waves.\cite{kol16}

The authors of ref.~\citenum{hum16} propose to use a low-frequency clock operating at the laser coherence limit to resolve the phase ambiguity in a second, high-frequency clock, which is interrogated near-synchronously and thus operates beyond the laser coherence limit.
This proposal exploits the dependence of the laser phase's standard deviation on frequency:
\begin{equation}
\Delta\phi_i \propto \nu_i T \sigma_y(T)
\end{equation}
with clock frequency $\nu_i$ where $i \in \{1,2\}$.
Similar to our protocol, it allows the low-frequency clock to provide a phase estimate to the high-frequency clock beyond the coherence time of the latter.
However, this protocol only considers two-pulse Ramsey interrogation of both clocks;
interrogation time is thus limited by the laser coherence time at the lower of the two clock frequencies.
The gains in interrogation time and in measurement variance are thus given by the clocks' frequency ratio.
Our protocol goes beyond this approach, as it allows interrogation beyond the coherence times of both clocks and works for arbitrary frequencies of the clocks.
Of course, a frequency ratio $\nu_2 / \nu_1 > 1$ of the two clocks can be exploited as discussed in ref.~\citenum{hum16} in addition to the partial decoupling.

Our scheme is compatible with the zero-dead-time scheme reported in ref.~\citenum{sch17}, which nearly eliminates the Dick effect and leads to QPN-limited frequency instability in optical lattice clocks.
A combined protocol can be implemented using four clocks.
As for single-ion clocks, our protocol improves frequency instability according to Eq.~(\ref{eq:sql}) and reduces the duration of measurements with a given precision by an order of magnitude or more in this case.

All these aspects make implementation of the compound protocol presented in this Letter highly appealing.
Interrogation times of one minute as supported by existing ultrastable lasers practically remove laser coherence as a limiting factor in most clocks.
They already allow resolving the natural line width in aluminium ion clocks, for instance.
Future ultrastable lasers will allow even longer interrogation times on the order of several minutes and beyond in compound clocks.
\begin{methods}
\subsection{Interrogation laser setup}
The interrogation laser of our strontium lattice clock at $\nu_\mathrm{Sr} \approx \SI{429}{\tera\hertz}$ has been described in detail in ref.~\citenum{hae15a}.
Its frequency stability has since been improved further by stability transfer from one of the ultrastable lasers described in ref.~\citenum{mat17a} via a single branch of an optical frequency comb using the transfer oscillator scheme.\cite{ste02a}
That laser is stabilised to a cryogenic monocrystalline silicon resonator near $\nu = \SI{194.4}{\tera\hertz}$.
It has a flicker frequency noise floor of about \num{4E-17} and similar white frequency noise at an averaging time $\tau = \SI{1}{\second}$ (see ref.~\citenum{mat17a}).

Laser light from the interrogation laser is delivered to the atoms via an optical fibre.
We have modified the optical path length stabilisation (PLS) such that it is compatible with multi-pulse interrogation (cf.\ ref.~\citenum{fal12}):
If a single beam is used for both spectroscopy and PLS  two subsequent pulses may differ in phase by $\Delta\varphi = \pi$ because the phase-locked loop (PLL) in the PLS is only sensitive to the round-trip phase.
Our observations have shown that these phase slips occur occasionally when the beam is switched on or off.
Therefore, we use two separate beams: a resonant probe beam for spectroscopy  and an off-resonant ($\Delta \nu = \SI{-2}{\mega\hertz}$) pilot beam, which remains switched on throughout spectroscopy, for the PLS servo loop.
Both beams are derived from the same acousto-optic modulator (AOM) in front of the fibre so that they nearly co-propagate.
The optical path length of the probe beam is then co-stabilised using the correction signal of the servo loop.
\subsection{Artificial frequency noise}
We imprint additional frequency noise directly onto the probe beam, by changing its radio frequency (rf) offset with respect to the pilot beam at the AOM of the PLS.
A direct digital synthesiser (DDS) based on a field programmable gate array (FPGA) provides an rf signal with the appropriate frequency noise for this purpose.
It uses pre-generated samples of pseudo-random coloured noise.\cite{kas95b}
Playback is controlled by the pulse pattern generator (PPG) of the clock apparatus for precise timing and reproducibility.

For the measurements reported in this Letter, we use artificial frequency noise spectra with $S_y(f) = \sum_{i=-1}^{0} h_i f^i$, where white frequency noise $h_0 = \SI{1.8E-31}{\per\hertz}$ (i.e., $\sigma_y = \num{3E-16}$ at $\tau = \SI{1}{\second}$) and flicker frequency noise $h_{-1} = \num{1.2E-30}$ ($\sigma_y = \num{1.3E-15}$) and $h_{-1} = \num{2.9E-30}$ ($\sigma_y = \num{2.0E-15}$), respectively, with respect to the transition frequency of the clock (see Fig.~\ref{fig:exp_scenario}a).
\subsection{Experimental setup} 
Our strontium optical lattice clock has been described in previous pub\-li\-ca\-tions.\cite{alm15, fal11}
For the experiments in this Letter, laser frequency noise is dominated by the artificial contribution discussed in the previous section and shown in Fig.~\ref{fig:exp_scenario}a.
We adjust the power of the spectroscopy beam such that it results in a $\pi$-pulse duration $T_\pi = \SI{1.0}{\milli\second}$, which keeps imperfections of the dynamic decoupling due to the Dick effect low.
A suitable spectroscopy sequence for clock 1 with an interrogation time $T_\mathrm{i}^\prime \approx \SI{495}{\milli\second}$ is determined following the procedure discussed in the main text.
We use a spectroscopy sequence with $M = 16$ such that $T_\mathrm{d} = \SI{30}{\milli\second} < T_\mathrm{co}$ and choose $\epsilon = 0.12 \pi$ such that the central fringe is broad enough to trace the expected phase deviations unambiguously.

The experiments reported in this Letter consist of phase reconstruction measurements of more than one thousand individual pre-generated noise samples.
The corresponding true phase deviations are calculated from the numerical noise samples by integrating the frequency noise over the interrogation time.

Phase reconstruction from the observed excitation probabilities is carried out using pre-generated look-up tables for the specific spectroscopy sequence given above, after correcting for imperfect contrast.
These measurements are interleaved with measurements without artificial noise, which are used to lock the laser frequency to the atomic transition frequency and compensate any frequency drift of the interrogation laser during the experiment.
We also observe and compensate for a light shift in clock 1.
However, such a shift does not impede operation of a compound clock or its systematic uncertainty, which is governed by clock 2.
\subsection{Phase noise due to near-synchronous interrogation}
Reading out the atomic state of clock 1 and forwarding the phase estimate to clock 2 (see Fig.~\ref{fig:protocol}b) introduces a brief window during which only one of the clocks is probed.
Laser phase noise that occurs during this time directly affects the determination of the correct Ramsey fringe in clock 2.
Hence, the delay $T_\mathrm{i}^\prime - T_\mathrm{i}$ must be kept short.
For the sake of simplicity, we use the same interrogation time in both clocks ($T_\mathrm{i} = T_\mathrm{i}^\prime$) for the experiments presented in this Letter, which use sequential interogation of a single clock (see main text).
In practice, the delay can easily be kept well below the laser coherence time of typical interrogation lasers.
\subsection{Phase noise due to dynamic decoupling imperfections}
Phase measurement noise in clock 1 caused by imperfections of the dynamic decoupling can be calculated analogously to the Dick effect,\cite{dic87} by analysis of the clock's sensitivity to laser frequency fluctuations.

The frequency sensitivity $g(t)$ of a spectroscopy protocol determines the change in excitation probability $\delta p_e$ caused by a time-dependent frequency error $\delta\nu(t)$ of the probe laser, which is given by\cite{dic87, dic90, que03}
\begin{equation}
\delta p_e = \frac{1}{2} \int_{0}^{T_\mathrm{i}} 2 \pi \delta\nu(t) g(t) \mathrm{d}t
\label{eq:dick_gt_basic}
\end{equation}
for interrogation time $T_\mathrm{i}$ in the linear response regime.
If the sensitivity function of clock 1 were identical to that of clock 2 except for a scaling factor there would be no imperfections of the dynamic decoupling.
In practice, this will never be the case due to the different pulse sequences used in the two clocks (see Fig.~\ref{fig:protocol}).
The frequency sensitivity $g(t)$ of clock 1 can be split into a signal component $\bar{g} = T_\mathrm{i}^{-1} \int_{0}^{T_\mathrm{i}} g(t)\mathrm{d}t$ and a noise component $g_\mathrm{n}(t) = g(t) - \bar{g}$ such that
\begin{equation}
\delta p_e = \frac{1}{2} \bar{g} \Delta\phi + \pi \int_{0}^{T_\mathrm{i}} g_\mathrm{n}(t) \delta\nu(t) \mathrm{d}t\mbox{,}
\label{eq:dick_gt_sn}
\end{equation}
where $\Delta\phi$ is the laser phase deviation accumulated during the interrogation time.
Laser frequency noise with single-sided PSD $S_y(f)$ thus gives rise to noise of the measured excitation probability with variance\cite{hob16a}
\begin{equation}
\sigma^2_{p_e} = \left(\pi \nu\right)^2 \int_{0}^{\infty} \left| \hat{g}_\mathrm{n}(f) \right|^2 S_y(f) \mathrm{d}f
\label{eq:dick_pe_var}
\end{equation}
where $\hat{g}_\mathrm{n}(f)$ is the complex Fourier transform of $g_\mathrm{n}(t)$.
This corresponds to phase measurement noise with variance
\begin{equation}
\sigma^2_{\Delta\phi} = \left( \frac{\bar{g}}{2} \right)^{-2} \sigma^2_{p_e}\mbox{.}
\label{eq:dick_phase_var}
\end{equation}

The frequency sensitivity of the protocol presented in this Letter is shown in Fig.~\ref{fig:exp_scenario}c for the parameters used in the proof-of-principle experiment.
The noise component $g_\mathrm{n}(t)$ stems mainly from the greatly increased frequency sensitivity during flip pulses.
Therefore, the phase measurement is particularly sensitive to laser frequency noise at Fourier frequencies that are harmonics of $f \approx (T_\pi + T_\mathrm{d})^{-1}$.
Sensitivity rolls off above a corner frequency $f_\mathrm{c} \approx T_\pi^{-1}$, which restricts useful $\pi$-pulse durations.

Noise due to these imperfections of the decoupling depends on the noise type ($S_y \propto f^\alpha$) that dominates at the relevant Fourier frequencies.
For flicker or white frequency noise ($\alpha = 0$ or $-1$), it decreases towards shorter pulse duration $T_\pi$, whereas it increases for flicker or white phase noise ($\alpha = 1$ or $2$).
White frequency noise is the dominant noise process for our laser system\cite{mat17a} at $T_\pi \approx \SI{1}{\milli\second}$.
The spectroscopy sequence used here gives rise to phase measurement noise with a standard deviation of \SI{0.39}{\radian} (\SI{0.04}{\radian}) from artificial (intrinsic) laser frequency noise.

In contrast with the procedure for estimating the frequency instability of atomic clocks\cite{dic87}, the dead time and total cycle time are not relevant for calculating the Dick effect in the dynamical decoupling scheme.
This is because the atomic excitation is being used in the dynamical decoupling scheme to estimate the phase accumulated by the laser only during the interrogation pulse, not the phase accumulated during the entire clock cycle.
\subsection{Phase noise due to QPN}
Like the dynamic decoupling imperfections, QPN causes phase measurement noise in clock 1 that may lead to incorrect determination of the Ramsey fringe.

The measured excitation probability has a variance $\sigma^2_{p_e} = \bar{p}_e (1 - \bar{p}_e) N^{-1}$ due to QPN for $N$ uncorrelated atoms, where $\bar{p}_e$ is the expectation value.
The variance of the resulting phase measurement noise is then given by Eq.~(\ref{eq:dick_phase_var}).
We use about \num{700} atoms for the experiments reported in this Letter.
This corresponds to $\sigma_{p_e} = 0.02$ and $\sigma_{\Delta\phi_\mathrm{est}} = \SI{0.2}{\radian}$.
\end{methods}
\section*{Bibliography}

\begin{thebibliography}{10}
	\expandafter\ifx\csname url\endcsname\relax
	\def\url#1{\texttt{#1}}\fi
	\expandafter\ifx\csname urlprefix\endcsname\relax\def\urlprefix{URL }\fi
	\providecommand{\bibinfo}[2]{#2}
	\providecommand{\eprint}[2][]{\url{#2}}
	
	\bibitem{san19}
	\bibinfo{author}{Sanner, C.} \emph{et~al.}
	\newblock \bibinfo{title}{Optical clock comparison test of {L}orentz symmetry}.
	\newblock \emph{\bibinfo{journal}{Nature}} \textbf{\bibinfo{volume}{567}},
	\bibinfo{pages}{204--208} (\bibinfo{year}{2019}).
	
	\bibitem{del17}
	\bibinfo{author}{Delva, P.} \emph{et~al.}
	\newblock \bibinfo{title}{Test of special relativity using a fiber network of
		optical clocks}.
	\newblock \emph{\bibinfo{journal}{Phys. Rev. Lett.}}
	\textbf{\bibinfo{volume}{118}}, \bibinfo{pages}{221102}
	(\bibinfo{year}{2017}).
	
	\bibitem{ros08}
	\bibinfo{author}{Rosenband, T.} \emph{et~al.}
	\newblock \bibinfo{title}{Frequency ratio of {Al$^+$} and {Hg$^+$} single-ion
		optical clocks; metrology at the 17th decimal place}.
	\newblock \emph{\bibinfo{journal}{Science}} \textbf{\bibinfo{volume}{319}},
	\bibinfo{pages}{1808--1812} (\bibinfo{year}{2008}).
	
	\bibitem{hun14}
	\bibinfo{author}{Huntemann, N.} \emph{et~al.}
	\newblock \bibinfo{title}{Improved limit on a temporal variation of $m_p / m_e$
		from comparisons of {Yb}$^+$ and {Cs} atomic clocks}.
	\newblock \emph{\bibinfo{journal}{Phys. Rev. Lett.}}
	\textbf{\bibinfo{volume}{113}}, \bibinfo{pages}{210802}
	(\bibinfo{year}{2014}).
	
	\bibitem{god14}
	\bibinfo{author}{Godun, R.~M.} \emph{et~al.}
	\newblock \bibinfo{title}{Frequency ratio of two optical clock transitions in
		$^{171}${Yb}$^+$ and constraints on the time-variation of fundamental
		constants}.
	\newblock \emph{\bibinfo{journal}{Phys. Rev. Lett.}}
	\textbf{\bibinfo{volume}{113}}, \bibinfo{pages}{210801}
	(\bibinfo{year}{2014}).
	
	\bibitem{der14}
	\bibinfo{author}{Derevianko, A.} \& \bibinfo{author}{Pospelov, M.}
	\newblock \bibinfo{title}{Hunting for topological dark matter with atomic
		clocks}.
	\newblock \emph{\bibinfo{journal}{Nature Physics}}
	\textbf{\bibinfo{volume}{10}}, \bibinfo{pages}{933--936}
	(\bibinfo{year}{2014}).
	
	\bibitem{wci16}
	\bibinfo{author}{Wcis{\l}o, P.} \emph{et~al.}
	\newblock \bibinfo{title}{Searching for topological defect dark matter with
		optical atomic clocks}.
	\newblock \emph{\bibinfo{journal}{Nat. Astron.}} \textbf{\bibinfo{volume}{1}},
	\bibinfo{pages}{0009} (\bibinfo{year}{2016}).
	
	\bibitem{sta16c}
	\bibinfo{author}{Stadnik, Y.~V.} \& \bibinfo{author}{Flambaum, V.~V.}
	\newblock \bibinfo{title}{Enhanced effects of variation of the fundamental
		constants in laser interferometers and application to dark-matter detection}.
	\newblock \emph{\bibinfo{journal}{Phys. Rev. A}} \textbf{\bibinfo{volume}{93}},
	\bibinfo{pages}{063630} (\bibinfo{year}{2016}).
	
	\bibitem{wci18a}
	\bibinfo{author}{Wcis{\l}o, P.} \emph{et~al.}
	\newblock \bibinfo{title}{New bounds on dark matter coupling from a global
		network of optical atomic clocks}.
	\newblock \emph{\bibinfo{journal}{Science Advances}}
	\textbf{\bibinfo{volume}{4}}, \bibinfo{pages}{eaau4869}
	(\bibinfo{year}{2018}).
	
	\bibitem{rob19b}
	\bibinfo{author}{Roberts, B.~M.} \emph{et~al.}
	\newblock \bibinfo{title}{Search for transient variations of the fine structure
		constant and dark matter using fiber-linked optical atomic clocks}.
	\newblock \bibinfo{howpublished}{arXiv:1907.02661 [astro-ph.CO]}
	(\bibinfo{year}{2019}).
	
	\bibitem{gro18a}
	\bibinfo{author}{Grotti, J.} \emph{et~al.}
	\newblock \bibinfo{title}{Geodesy and metrology with a transportable optical
		clock}.
	\newblock \emph{\bibinfo{journal}{Nature Physics}}
	\textbf{\bibinfo{volume}{14}}, \bibinfo{pages}{437--441}
	(\bibinfo{year}{2018}).
	
	\bibitem{kol16}
	\bibinfo{author}{Kolkowitz, S.} \emph{et~al.}
	\newblock \bibinfo{title}{Gravitational wave detection with optical lattice
		atomic clocks}.
	\newblock \emph{\bibinfo{journal}{Phys. Rev. D}} \textbf{\bibinfo{volume}{94}},
	\bibinfo{pages}{124043} (\bibinfo{year}{2016}).
	
	\bibitem{hae15a}
	\bibinfo{author}{H{\"a}fner, S.} \emph{et~al.}
	\newblock \bibinfo{title}{8 $ \times 10^{-17}$ fractional laser frequency
		instability with a long room-temperature cavity}.
	\newblock \emph{\bibinfo{journal}{Opt. Lett.}} \textbf{\bibinfo{volume}{40}},
	\bibinfo{pages}{2112--2115} (\bibinfo{year}{2015}).
	
	\bibitem{mat17a}
	\bibinfo{author}{Matei, D.~G.} \emph{et~al.}
	\newblock \bibinfo{title}{$1.5~\mu$m lasers with sub-{10 mHz} linewidth}.
	\newblock \emph{\bibinfo{journal}{Phys. Rev. Lett.}}
	\textbf{\bibinfo{volume}{118}}, \bibinfo{pages}{263202}
	(\bibinfo{year}{2017}).
	
	\bibitem{cho11}
	\bibinfo{author}{Chou, C.~W.}, \bibinfo{author}{Hume, D.~B.},
	\bibinfo{author}{Thorpe, M.~J.}, \bibinfo{author}{Wineland, D.~J.} \&
	\bibinfo{author}{Rosenband, T.}
	\newblock \bibinfo{title}{Quantum coherence between two atoms beyond
		$q=10^{15}$}.
	\newblock \emph{\bibinfo{journal}{Phys. Rev. Lett.}}
	\textbf{\bibinfo{volume}{106}}, \bibinfo{pages}{160801}
	(\bibinfo{year}{2011}).
	
	\bibitem{tak11}
	\bibinfo{author}{Takamoto, M.}, \bibinfo{author}{Takano, T.} \&
	\bibinfo{author}{Katori, H.}
	\newblock \bibinfo{title}{Frequency comparison of optical lattice clocks beyond
		the {D}ick limit}.
	\newblock \emph{\bibinfo{journal}{Nature Photonics}}
	\textbf{\bibinfo{volume}{5}}, \bibinfo{pages}{288--292}
	(\bibinfo{year}{2011}).
	
	\bibitem{hum16}
	\bibinfo{author}{Hume, D.~B.} \& \bibinfo{author}{Leibrandt, D.~R.}
	\newblock \bibinfo{title}{Probing beyond the laser coherence time in optical
		clock comparisons}.
	\newblock \emph{\bibinfo{journal}{Phys. Rev. A}} \textbf{\bibinfo{volume}{93}},
	\bibinfo{pages}{032138} (\bibinfo{year}{2016}).
	
	\bibitem{sch17}
	\bibinfo{author}{Schioppo, M.} \emph{et~al.}
	\newblock \bibinfo{title}{Ultra-stable optical clock with two cold-atom
		ensembles}.
	\newblock \emph{\bibinfo{journal}{Nature Photonics}}
	\textbf{\bibinfo{volume}{11}}, \bibinfo{pages}{48--52}
	(\bibinfo{year}{2017}).
	
	\bibitem{oel19}
	\bibinfo{author}{Oelker, E.} \emph{et~al.}
	\newblock \bibinfo{title}{Demonstration of $4.8\times10^{-17}$ stability at 1 s
		for two independent optical clocks}.
	\newblock \emph{\bibinfo{journal}{Nature Photonics}}
	\textbf{\bibinfo{volume}{13}}, \bibinfo{pages}{714--719}
	(\bibinfo{year}{2019}).
	
	\bibitem{ita93}
	\bibinfo{author}{Itano, W.~M.} \emph{et~al.}
	\newblock \bibinfo{title}{Quantum projection noise: Population fluctuations in
		two-level systems}.
	\newblock \emph{\bibinfo{journal}{Phys. Rev. A}} \textbf{\bibinfo{volume}{47}},
	\bibinfo{pages}{3554--3570} (\bibinfo{year}{1993}).
	\newblock \bibinfo{note}{See Also: Erratum Phys. Rev. A 51, 1717 (1995)}.
	
	\bibitem{dic87}
	\bibinfo{author}{Dick, G.~J.}
	\newblock \bibinfo{title}{Local oscillator induced instabilities in trapped ion
		frequency standards}.
	\newblock In \emph{\bibinfo{booktitle}{Proceedings of $19^{th}$ Annu. Precise
			Time and Time Interval Meeting, Redendo Beach, 1987}},
	\bibinfo{pages}{133--147} (\bibinfo{publisher}{U.S. Naval Observatory},
	\bibinfo{address}{Washington, DC}, \bibinfo{year}{1988}).
	\newblock
	\urlprefix\url{http://tycho.usno.navy.mil/ptti/1987papers/Vol\%2019_13.pdf}.
		
		\bibitem{que03}
		\bibinfo{author}{Quessada, A.} \emph{et~al.}
		\newblock \bibinfo{title}{The {D}ick effect for an optical frequency standard}.
		\newblock \emph{\bibinfo{journal}{J. Opt. B: Quantum Semiclass. Opt.}}
		\textbf{\bibinfo{volume}{5}}, \bibinfo{pages}{S150--S154}
		(\bibinfo{year}{2003}).
		
		\bibitem{rie04a}
		\bibinfo{author}{Riehle, F.}
		\newblock \emph{\bibinfo{title}{Frequency Standards: Basics and Applications}}
		(\bibinfo{publisher}{Wiley-VCH}, \bibinfo{address}{Weinheim},
		\bibinfo{year}{2004}).
		
		\bibitem{win92}
		\bibinfo{author}{Wineland, D.~J.}, \bibinfo{author}{Bollinger, J.~J.},
		\bibinfo{author}{Itano, W.~M.}, \bibinfo{author}{Moore, F.~L.} \&
		\bibinfo{author}{Heinzen, D.~J.}
		\newblock \bibinfo{title}{Spin squeezing and reduced quantum noise in
			spectroscopy}.
		\newblock \emph{\bibinfo{journal}{Phys. Rev. A}} \textbf{\bibinfo{volume}{46}},
		\bibinfo{pages}{R6797--R6800} (\bibinfo{year}{1992}).
		
		\bibitem{ros07}
		\bibinfo{author}{Rosenband, T.} \emph{et~al.}
		\newblock \bibinfo{title}{Observation of the {$^1$S$_0$} - {$^3$P$_0$} clock
			transition in {$^{27}$Al$^+$}}.
		\newblock \emph{\bibinfo{journal}{Phys. Rev. Lett.}}
		\textbf{\bibinfo{volume}{98}}, \bibinfo{pages}{220801}
		(\bibinfo{year}{2007}).
		
		\bibitem{rob97}
		\bibinfo{author}{Roberts, M.} \emph{et~al.}
		\newblock \bibinfo{title}{Observation of an electric octupole transition in a
			single ion}.
		\newblock \emph{\bibinfo{journal}{Phys. Rev. Lett.}}
		\textbf{\bibinfo{volume}{78}}, \bibinfo{pages}{1876--1879}
		(\bibinfo{year}{1997}).
		
		\bibitem{vio99}
		\bibinfo{author}{Viola, L.}, \bibinfo{author}{Knill, E.} \&
		\bibinfo{author}{Lloyd, S.}
		\newblock \bibinfo{title}{Dynamical decoupling of open quantum systems}.
		\newblock \emph{\bibinfo{journal}{Phys. Rev. Lett.}}
		\textbf{\bibinfo{volume}{82}}, \bibinfo{pages}{2417--2421}
		(\bibinfo{year}{1999}).
		
		\bibitem{hah50}
		\bibinfo{author}{Hahn, E.~L.}
		\newblock \bibinfo{title}{Spin echos}.
		\newblock \emph{\bibinfo{journal}{Phys. Rev.}} \textbf{\bibinfo{volume}{80}},
		\bibinfo{pages}{580--594} (\bibinfo{year}{1950}).
		
		\bibitem{gre16}
		\bibinfo{author}{Grebing, C.} \emph{et~al.}
		\newblock \bibinfo{title}{Realization of a timescale with an accurate optical
			lattice clock}.
		\newblock \emph{\bibinfo{journal}{Optica}} \textbf{\bibinfo{volume}{3}},
		\bibinfo{pages}{563--569} (\bibinfo{year}{2016}).
		
		\bibitem{ste02a}
		\bibinfo{author}{Stenger, J.}, \bibinfo{author}{Schnatz, H.},
		\bibinfo{author}{Tamm, C.} \& \bibinfo{author}{Telle, H.~R.}
		\newblock \bibinfo{title}{Ultra-precise measurement of optical frequency
			ratios}.
		\newblock \emph{\bibinfo{journal}{Phys. Rev. Lett.}}
		\textbf{\bibinfo{volume}{88}}, \bibinfo{pages}{073601}
		(\bibinfo{year}{2002}).
		
		\bibitem{fal12}
		\bibinfo{author}{Falke, S.}, \bibinfo{author}{Misera, M.},
		\bibinfo{author}{Sterr, U.} \& \bibinfo{author}{Lisdat, C.}
		\newblock \bibinfo{title}{Delivering pulsed and phase stable light to atoms of
			an optical clock}.
		\newblock \emph{\bibinfo{journal}{Appl. Phys. B}}
		\textbf{\bibinfo{volume}{107}}, \bibinfo{pages}{301--311}
		(\bibinfo{year}{2012}).
		
		\bibitem{kas95b}
		\bibinfo{author}{Kasdin, N.~J.}
		\newblock \bibinfo{title}{Discrete simulation of colored noise and stochastic
			processes and $1/f^{\alpha}$ power law noise generation}.
		\newblock \emph{\bibinfo{journal}{Proc. IEEE}} \textbf{\bibinfo{volume}{83}},
		\bibinfo{pages}{802--827} (\bibinfo{year}{1995}).
		
		\bibitem{alm15}
		\bibinfo{author}{Al-Masoudi, A.}, \bibinfo{author}{D\"orscher, S.},
		\bibinfo{author}{H\"afner, S.}, \bibinfo{author}{Sterr, U.} \&
		\bibinfo{author}{Lisdat, C.}
		\newblock \bibinfo{title}{Noise and instability of an optical lattice clock}.
		\newblock \emph{\bibinfo{journal}{Phys. Rev. A}} \textbf{\bibinfo{volume}{92}},
		\bibinfo{pages}{063814} (\bibinfo{year}{2015}).
		
		\bibitem{fal11}
		\bibinfo{author}{Falke, S.} \emph{et~al.}
		\newblock \bibinfo{title}{The $^{87}${S}r optical frequency standard at {PTB}}.
		\newblock \emph{\bibinfo{journal}{Metrologia}} \textbf{\bibinfo{volume}{48}},
		\bibinfo{pages}{399--407} (\bibinfo{year}{2011}).
		
		\bibitem{dic90}
		\bibinfo{author}{Dick, G.~J.}, \bibinfo{author}{Prestage, J.},
		\bibinfo{author}{Greenhall, C.} \& \bibinfo{author}{Maleki, L.}
		\newblock \bibinfo{title}{Local oscillator induced degradation of medium-term
			stability in passive atomic frequency standards}.
		\newblock In \emph{\bibinfo{booktitle}{Proceedings of the 22nd Annual Precise
				Time and Time Interval (PTTI) Applications and Planning Meeting, Vienna VA,
				USA}}, \bibinfo{pages}{487--509} (\bibinfo{year}{1990}).
		\newblock \urlprefix\url{http://tycho.usno.navy.mil/ptti/1990/Vol\%2022_42.pdf}.
			
			\bibitem{hob16a}
			\bibinfo{author}{Hobson, R.}
			\newblock \emph{\bibinfo{title}{An optical lattice clock with neutral
					strontium}}.
			\newblock Ph.D. thesis, \bibinfo{school}{Balliol College, University of Oxford}
			(\bibinfo{year}{2016}).
			\newblock
			\urlprefix\url{https://ora.ox.ac.uk/objects/uuid:d52faaaf-307c-4b48-847f-be590f46136f}.
			
		\end{thebibliography}

%
%
%
\begin{addendum}
	\item We thank S.\ Häfner, Th.\ Legero, and E.\ Benkler for operating the ultrastable laser system and optical frequency comb that are used for stabilisation of our interrogation laser system.
	We thank M.\ Misera for developing the frequency noise generator.
	This work has received funding from Deutsche Forschungsgemeinschaft (DFG, German Research Foundation) within CRC 1227 (``DQ-mat'', project B02) and under Germany’s Excellence Strategy –- EXC-2123/1 (``QuantumFrontiers'').
	R.H.\ has been supported by the European Union's Horizon H2020 MSCA RISE program under Grant Agreement Number 691156 (Q-SENSE).
	M.B.\ has been supported by a research fellowship within the project ``Enhancing Educational Potential of Nicolaus Copernicus University in the Disciplines of Mathematical and Natural Sciences'' (project no.\ POKL.04.01.01-00-081/10).
	U.S. acknowledges funding from the project EMPIR 17FUN03 USOQS.
	EMPIR projects are co-funded by the European Union’s Horizon2020 research and innovation programme and the EMPIR Participating States.
	\item[Contributions] S.D., M.B., R.H., U.S., and C.L.\ developed the protocol and devised the proof-of-principle experiement;
	S.D., A.A., M.B., R.S., R.H., and C.L.\ fitted the strontium clock for the experiment;
	S.D., A.A., and R.S.\ acquired and analysed the data.
	All authors were involved in discussions and preparation of the manuscript.
	\item[Competing interests] The authors declare no competing interests.
	\item[Correspondence]\begin{flushleft}
		Correspondence and requests for materials should be addressed to S.D.\ \linebreak[4](email: \mbox{soeren.doerscher@ptb.de}) or C.L.\ (email: \mbox{christian.lisdat@ptb.de}).
	\end{flushleft} 
\end{addendum}
\end{document}